# Data Processing Chain and Products of EOS-06 OCM-3 Payload

## From Signal Processing to Geometric Precision

Ankur Garg, Space Applications Centre **(Corresponding Author),**
**Contact : agarg@sac.isro.gov.in**

Tushar Shukla, Space Applications Centre
Sunita Arya, Space Applications Centre
Ghansham Sangar, Space Applications Centre
Meenakshi Sarkar, Space Applications Centre
Sampa Roy, Space Applications Centre
S. Manthira Moorthi, Space Applications Centre
Debajyoti Dhar, Space Applications Centre

## Acknowledgment

The authors thankfully acknowledges the understanding, encouragement and support received from Director, Space Applications Centre, ISRO. The authors would also like to thank other Signal and Image Processing Area (SIPA) members who have given their support from time to time. The continuing support from IRS project Management through IRS Program Director, Oceansat-3 Project Director, Associate Project Director, Payloads, Applications and Ground Segment Committee members is thankfully acknowledged.

## Funding N/A

## Declarations

**Conflict of Interest** The authors declared that they have no conflict of interest.

**Abstract.** The Ocean Color Monitor-3 (OCM-3), launched aboard Oceansat-3, represents a significant advancement in ocean observation technology, building upon the capabilities of its predecessors. With thirteen spectral bands, OCM-3 enhances feature identification and atmospheric correction, enabling precise data collection from a sun-synchronous orbit. Operating at an altitude of 732.5 km, the satellite achieves high signal-to-noise ratios (SNR) through sophisticated onboard and ground processing techniques, including advanced geometric modeling for pixel registration. The OCM-3 processing pipeline, consisting of multiple levels, ensures rigorous calibration and correction of radiometric and geometric data. This paper presents key methodologies such as dark data modeling, photo response non-uniformity correction, and smear correction, are employed to enhance data quality. The effective implementation of ground time delay integration (TDI) allows for the refinement of SNR, with evaluations demonstrating that performance specifications were exceeded. Geometric calibration procedures, including band-to-band registration and geolocation accuracy assessments, which further optimize data reliability are presented in the paper. Advanced image registration techniques leveraging Ground Control Points (GCPs) and residual error analysis significantly reduce geolocation errors, achieving precision within specified thresholds. Overall, OCM-3's comprehensive calibration and processing strategies ensure high-quality, reliable data crucial for ocean monitoring and change detection applications, facilitating improved understanding of ocean dynamics and environmental changes.

## 1 Introduction

The Ocean Color Monitor-3 (OCM-3), which launched aboard Oceansat-3, continues the legacy of Oceansat-1 and Oceansat-2 by providing essential data services for ocean applications. OCM-3 features thirteen spectral bands, an enhancement over the eight bands of its predecessors, which significantly improves feature identification and atmospheric correction in the visible and near-infrared regions of the electromagnetic spectrum.

Oceansat-3 operates in a sun-synchronous orbit at an altitude of 732.5 km with an inclination of 98.32 degrees. This orbit allowes the satellite to observe the same target from six different angles over a 13-day period, after which the ground track repeated. The mission aims to achieve a high signal-to-noise ratio (SNR) at ocean reference radiances through onboard binning and ground-based time delay and integration processing [1]. Data is imaged at a resolution of 366 meters in Local Area Coverage (LAC) mode and approximately 1.1 kilometers in Global Area Coverage (GAC) mode, with a swath width of around 1590 km, as detailed in Table 1 including major payload parameters.

**Table 1.** OCM-3 Payload Parameters

| S. No. | Parameter | Value | Remarks |
|---|---|---|---|
| 1. | Resolution (m) | 366/732(AX) x 1100(AL) | LAC/GAC @Nadir $0^0$ tilt |
| 2. | Swath (km) | ~1590 | |
| 3. | Detector Pixel Pitch (um) | 10 | |
| 4. | Focal Length (mm) | 20 | |
| 5. | Field of View (deg) | ± 43.5 | |
| 6. | Frame Size (active pixels) | 4000 (AX) x 48 (AL) | |
| 7. | Oversampling Factor within GSD | 1-8 | Varies across band |
| 8. | Integration Time/Frame (ms) | 7.98-65.8 | Varies across band, based on Oversampling and altitude |
| 9. | Radiometric Resolution (bits) | 12/16 | LAC/GAC |
| 10. | Number of Channels | 13 | |
| 11. | Compression Ratio | 1.8 | |
| 12. | SNR Goal at Ocean Reference Radiance | >=800 for B1-B10 (@366m) >=600 for B11-B13 (@1.1km) | |
| 13. | Band To Band Registration (pixels) | ±1 | |

OCM-3 operates at an altitude of 734-767 km with an inclination of 98.331 degrees, as shown in Table 2 along with other major orbital parameters. This orbital configuration allows for effective data collection, ground track repeatability and path to path shift to avoid sun glint area.

**Table 2.** EOS-06 Orbital Parameters

| S. No. | Parameter | Value |
|---|---|---|
| 1. | Altitude (km) | 734-767 |
| 2. | Inclination(deg) | 98.331 |
| 3. | Eccentricity | 0.0012 |
| 4. | Local Time (hours) | 12:00 |
| 5. | Q Factor | 14 6/13 |
| 6. | Ground Path Repeat(days) | 13 |
| 7. | No. of orbits for ground track repeat | 188 |
| 8. | Path to Path separation(km) | 213 |
| 9. | No. of paths covered by payload | 6 |
| 10. | No. of days for coverage | 2 |
| 11. | No. days for glint free product | 4/5 |

The mission specifications outlined in Table 3 indicates a pointing accuracy of less than 600 meters and a location accuracy of less than 500 meters. The specification of drift rate is critical for frame to frame alignment and effective Ground TDI working.

**Table 3.** OCM-3 Mission Specifications

| S. No. | Parameter | Value |
|---|---|---|
| 1. | Pointing Accuracy | <600m |
| 2. | Satellite Drift | 6e-4 (deg/sec)<br>5e-4 (deg/sec) (With SS Update) |
| 3. | Jitter | 1.53e-3 deg |
| 4. | Location Accuracy | <500m |

The OCM-3 employes an advanced imaging scheme that enabled it to achieve a signal-to-noise ratio (SNR) exceeding requirements, utilizing both in-orbit and ground processing binning methods. The satellite’s frame camera is equipped with 4000 across-track pixels and 48 along-track pixels, tilted at ±20 degrees to minimize sun glint effects as shown in Figure 1. The camera captures frames over 64 milliseconds imaging the same ground location 48 different time. Consecutive farmes are biinned onboard generating 24 LAC samples for each ground feature [1] which are then transmitted to the ground for further processing.

The large field of view (FOV) of ±43.5 degrees introduces distortions, leading to variations in the instantaneous ground field of view (IGFOV) across different rows. Consequently, the same ground object could be represented across multiple pixels, with sub-pixel shifts resulting from satellite motion and optical distortions as shown in Figure 2. To address these challenges, OCM-3 on ground processing utilizes a well-calibrated geometric model that established precise relationships between pixels across frames, accurately mapping image data to corresponding ground positions while accounting for motion and distortions. Knowledge of the satellite's attitude—roll, pitch, and yaw—is crucial for correcting these distortions and ensuring precise pixel registration. Although onboard binning enhances data management efficiency, it complicated pixel-to-ground correspondence. The geometric model effectively navigated these complexities, facilitating accurate spatial representation essential for reliable Earth observation.

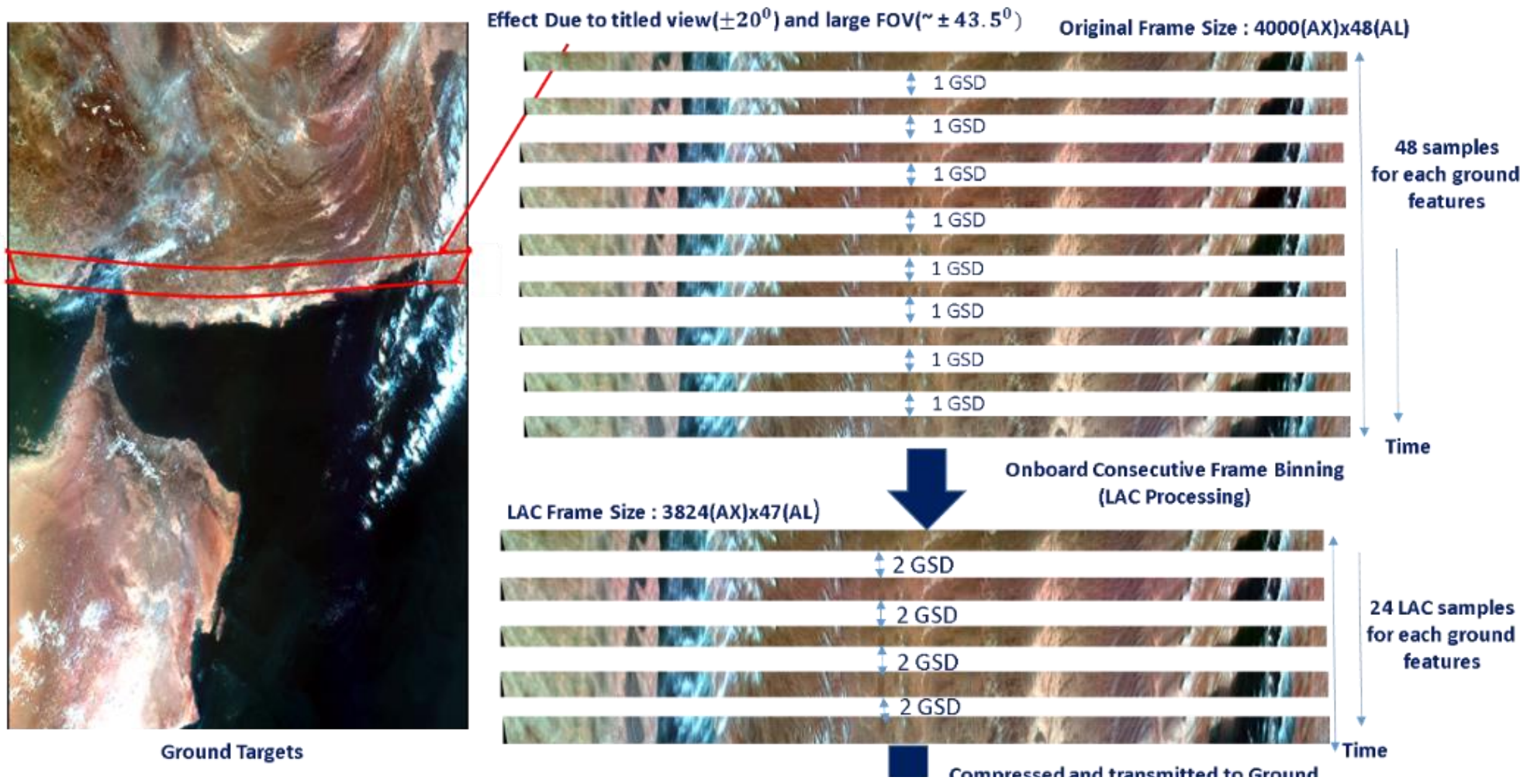

**Figure 1.** OCM-3 Imaging Scheme

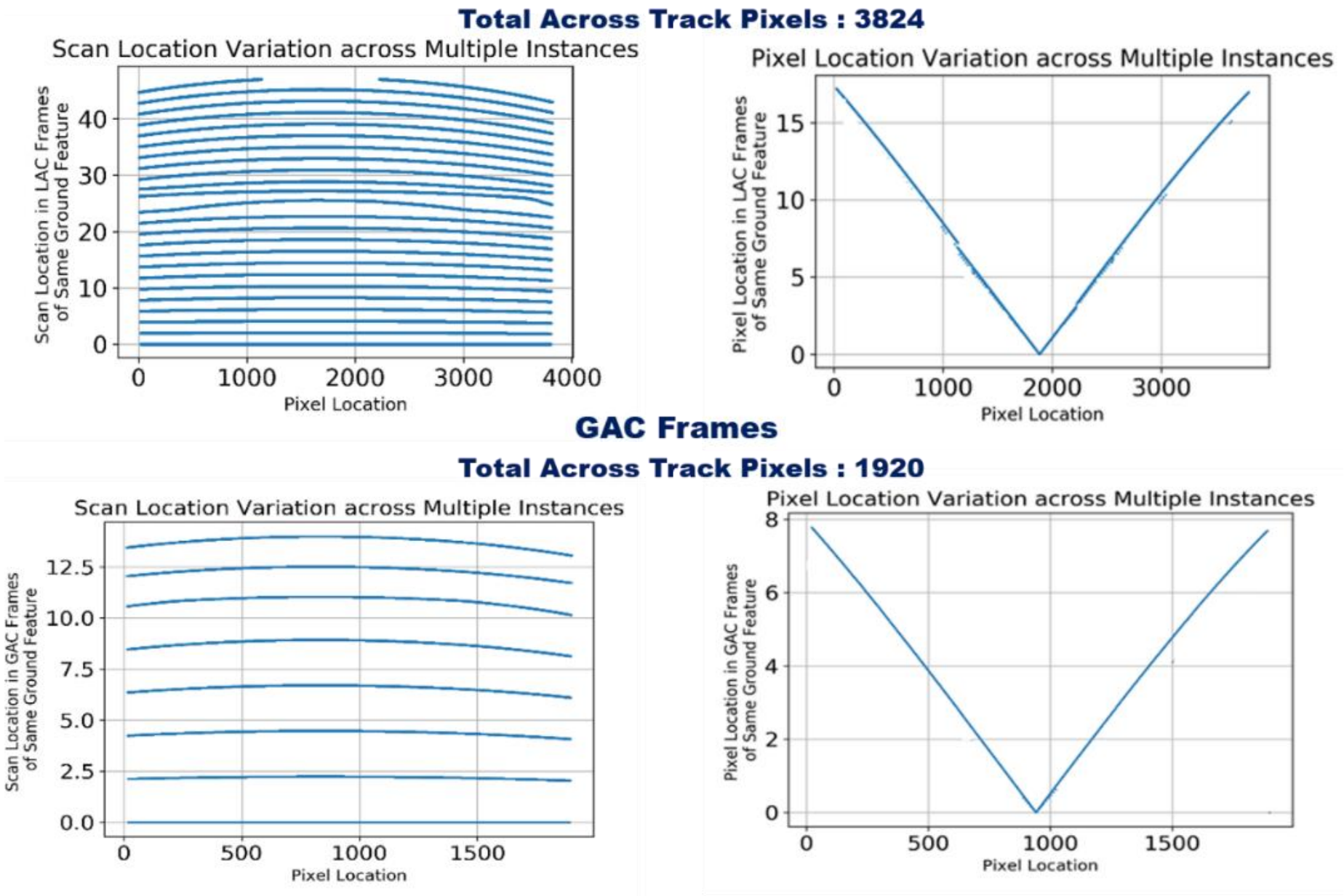

**Figure 2.** Distribution of Ground Location in the OCM-3 Frame Array

## 2 Data Products Overview

The primary objective of OCM-3 data processing is to generate high SNR images from both LAC and GAC frames while meeting radiometric and geometric accuracy requirements. Table 4 outlined the key product requirements, emphasizing that the specification for band-to-band registration (BBR) must remain within ±0.25 pixels, which was critical for ensuring data integrity.

**Table 4.** OCM-3 Product Requirements

| S. No. | Parameter | Value(CE90) |
|---|---|---|
| 1. | Band to Band Registration | ±0.25 pixels |
| 2. | Location Accuracy | ±1 km |
| 3. | Internal Distortion | ±2 pixels |

Table 5 illustrates the data processing levels for OCM-3, highlighting the introduction of Level-1C and Level-2C for GAC mode to ensure compatibility with SSTM-1 data products for the generation of higher-level products such as PFZ as compared to OCM-1 and OCM-2 which only offered Level-1B data.

**Table 5.** Data Processing Levels

| S. No. | Level | Description | Remarks | Mode |
|---|---|---|---|---|
| 1 | Level-1 Basic Data Products | Level-1B: Radiance Product (Radiometrically Corrected and Geotagged) | | LAC & GAC |
| | | Level-1C : Geo-referenced Product (Geo-rectified Map Projected) | Ellipsoid : WGS-84 Map Projection : LCC | LAC & GAC |
| 2 | Level-2 Geophysical Parameters | Level-2B: Geo-physical parameters (Radiometrically Corrected and Geotagged) | | LAC & GAC |
| | | Level-2C: Geo-physical parameters (Geo-rectified Map Projected ) | Ellipsoid : WGS-84 Map Projection : LCC | LAC & GAC |

## 3 OCM-3 Processing Chain

The data processing chain for OCM-3 is shown in Figure 3. The initial processing step, known as Level-0, includes demodulation, decryption, and decompression of the space packets received on the ground. Following Level-0 processing, frame-wise radiometric corrections is applied, encompassing dark data modeling, pixel response non-uniformity correction, and frame transfer smear correction. The radiometrically corrected frames were then processed through the ground time delay integration (TDI) algorithm, which integrates digital elevation models (DEMs), alignment angles, orbit and attitude information, and band to band mis registration (BBR) profiles to generate Level-1 products.

The Level-1 outputs are subsequently fed into an atmospheric correction process that handles land and ocean regions separately [2]. These results are used in the geo-physical parameter estimation chain, ultimately leading to the generation of oceanic and terrestrial geophysical products. This comprehensive processing sequence en-sured that OCM-3 data was accurately calibrated and ready for further analysis and application.

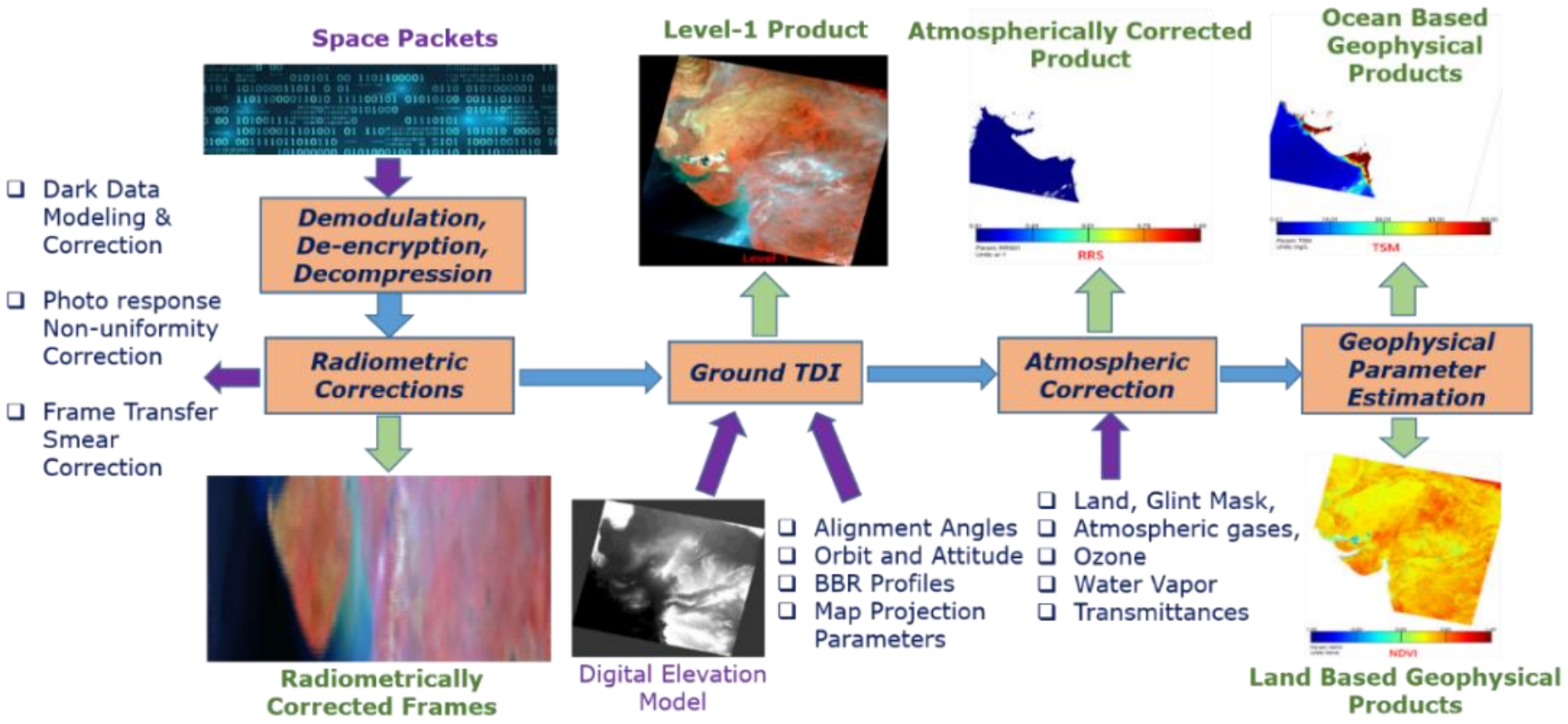

**Figure 3.** OCM-3 Ground Processing Chain

The process for computing the Top of Atmosphere (TOA) Earth radiance from the raw Earth signal received from Level-0 involves a detailed model that operates on a per-pixel basis [3]. The equation used for this calculation can be expressed as follows:

$$L_{calib}\,(k,i,j) = Nonlin_{k,i,j}\,(V(k,i,j)-DS(k,i)) * C_{k,i} + D_{k,i}$$

In the above equation,
$L_{Calib}(k,i,j)$ : Radiometrically corrected Earth science data (TOA radiance)
$Nonlin_{k,i,j}(X)$ : Linearization function for each pixel $(k,i,j)$
V(k,i,j) : Raw Earth data
$C_{k,i}$, $D_{k,i}$: Ground-calibration multiplicative and additive coefficients for count to radiance conversion
where k is the band index, i is the row index and j is the column index of the frame

**Dark Modeling and Correction :** The OCM-3 payload incorporates a shielded row to provide dark response data across all 3824 columns, enabling effective correction for both Local Area Coverage (LAC) and Global Area Coverage (GAC) modes. Specifically, one row of data is transmitted every four WLS for LAC mode and every twelve WLS for GAC mode. Just after the launch, on November 27, 2021, a night imaging session was conducted in LAC mode using Strip ID 32. A comparative analysis of the night imaging data and the corresponding dark row data, depicted in Figure 2, reveals a close correlation, with discrepancies of only 3-4 counts between the actual dark data and the recorded dark row data. This close match suggests that the dark row data could be reliably utilized to correct for in-orbit data anomalies. However, due to the inherent noise in the data received from this shielded row, a method was imple-mented to scale the column-wise mean profile derived from the night imaging data with the current row data, facilitating effective correction of port biases. This approach enhanced the overall accuracy and reliability of the image processing. Figure 4 compares the product quality before and after applying port-wise dark correction, showcasing the improvements achieved in LAC Image. The port biases seen in the image before dark correction using row data have got corrected.

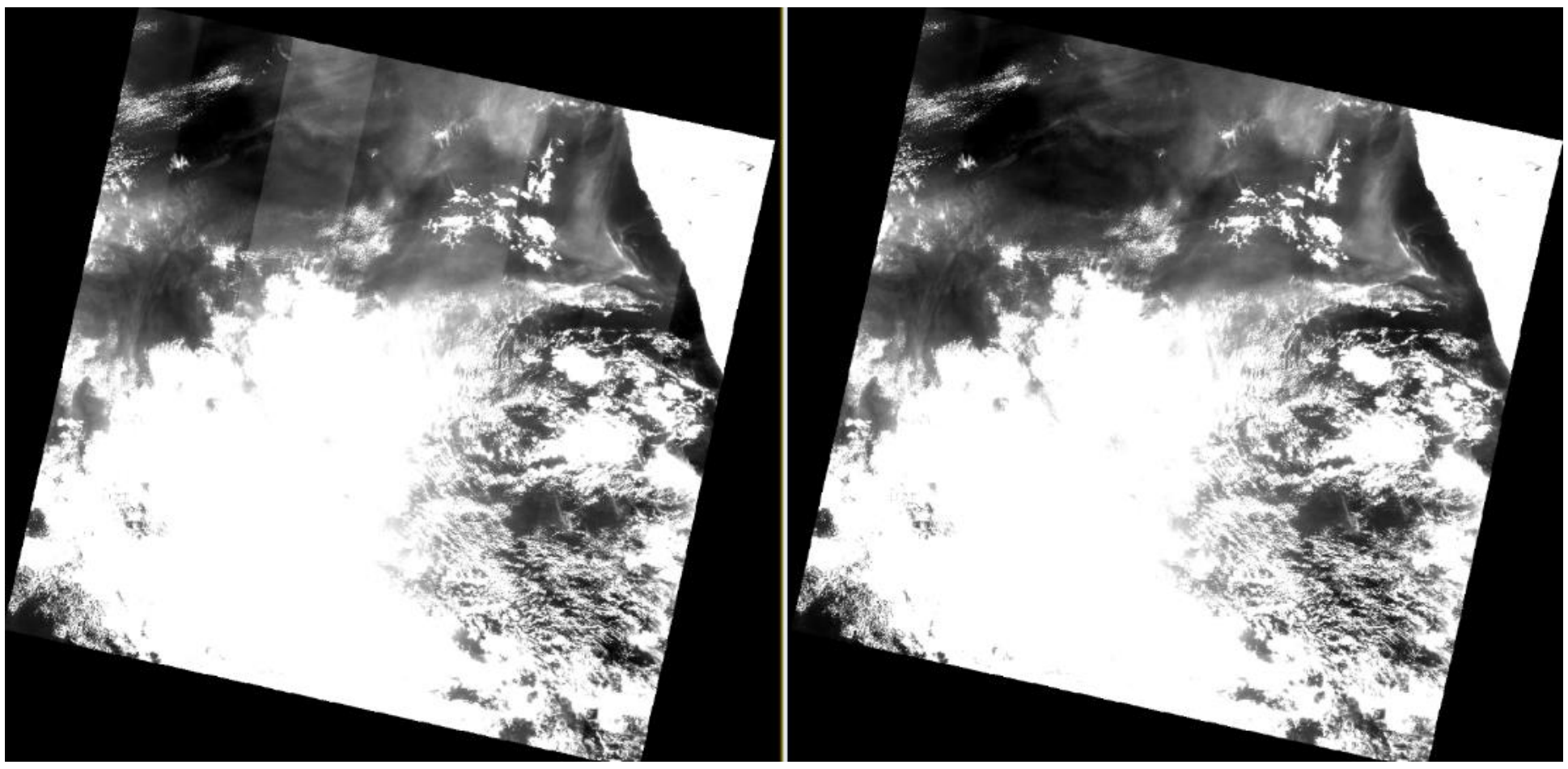

**Figure 4.** Band-12 Before and After Dark Correction

**Photo Response Non-Uniformity Correction:** Photo Response Non-Uniformity (PRNU) refers to the variations in sensitivity among individual pixels of a sensor, leading to inconsistencies in the response to uniform illumination. These variations can arise from several factors, including manufacturing discrepancies, variations in pixel architecture, and changes in operational conditions, such as temperature fluctuations and aging of the sensor components. PRNU can significantly affect the accuracy of remotely sensed data, as it can introduce artifacts and distortions in the captured images. To address PRNU, pre-computed information generated from both laboratory and in-orbit calibration is being utilized during data processing. This calibration helps identify and characterize the non-uniformities present in the sensor’s response. To refine the PRNU corrections further, multiple datasets are collected across various terrains and conditions inorbit. With central limit theorem, it is assumed that all pixels have seen all landscapes and should respond similarly when exposed to the same dataset as these landscapes simulate uniform illumination conditions. By analyzing the differences observed across these datasets, it becomes possible to isolate variations attributed specifically to PRNU. These differences are then employed to calculate relative gains for each pixel, which normalizes the response across the entire field of view (FOV) [4]. This method ensures that the sensor's output becomes more uniform, enhancing the overall quality and reliability of the data captured by the satellite. Figure 5 shows the result of in-orbit PRNU before and after corrections. The strips visible in original image is corrected in the corrected image.

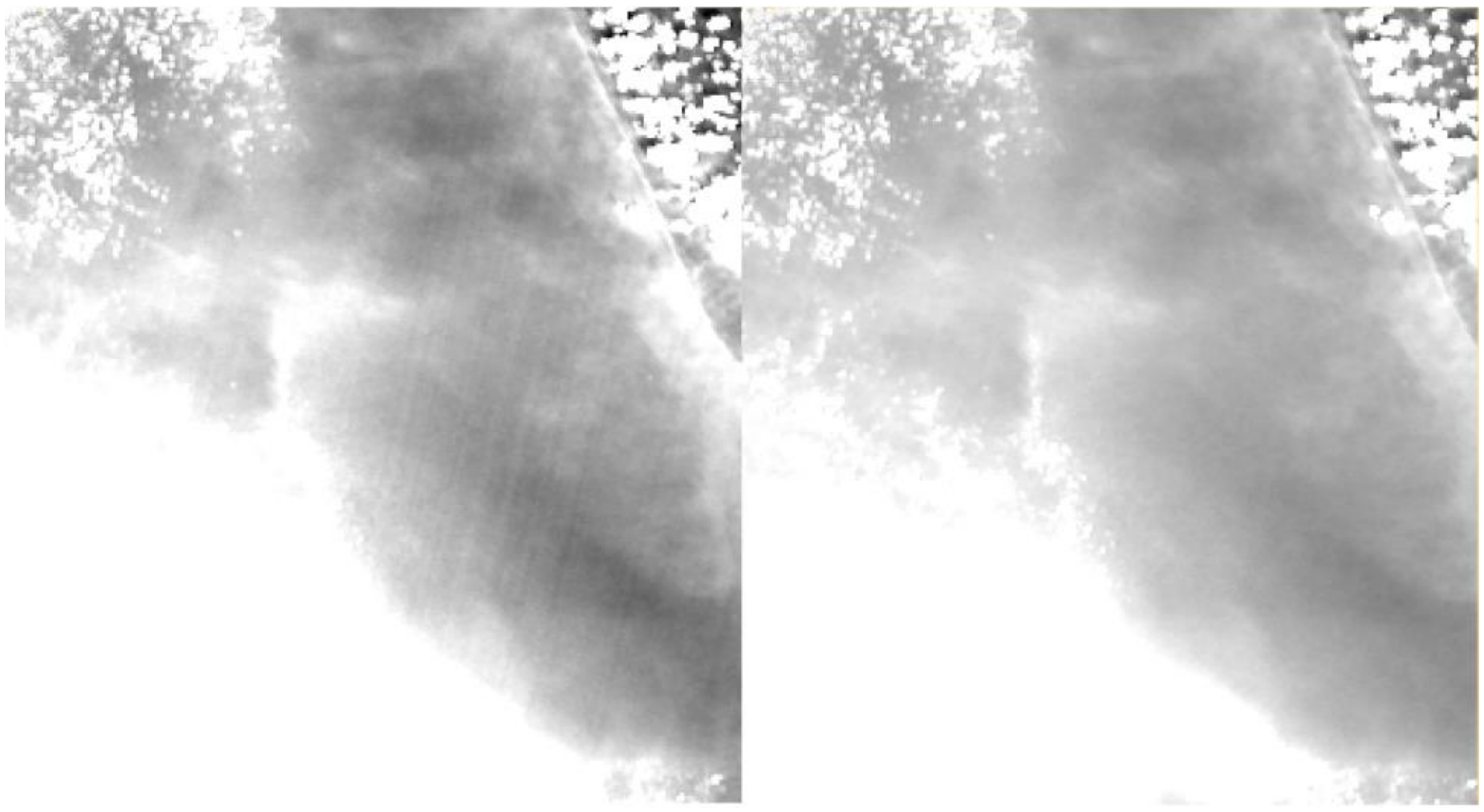

**Figure 5.** LAC Band-1 Before and After PRNU Coefficients Updation

**Smear Correction:** OCM-3 data exhibits smear artifacts resulting from the frame transfer-based image acquisition methodology [5]. The transfer time for each row from the image section to the storage section is 2 microseconds, during which data acquisition continues. Given that the detector comprises a total of 54 rows, the complete transfer time for all rows sums up to 108 microseconds. Since data is not flushed from the image section after transferring one complete frame, residual signals from the previous frame can influence the current frame, resulting in bidirectional smear. Consequently, each LAC/GAC frame is affected by two ground features from the preceding frame and one from the subsequent frame. Due to the onboard binning processes employed to generate LAC and GAC frames, the true values of missing samples are not directly available; instead, they are estimated from the current frame. To address this, a weights matrix has been developed to account for the charge collection and transfer scheme based on LAC/GAC mode of acquisition and with/without exposure time considering the oversampling and variable integration time setting. Each frame is corrected by inverting this weight matrix, which mitigates the smear effects and enhances data accuracy. This smear is context-dependent and can lead to variability in spectral signatures across different dates, as illustrated in Figure 5 for the MOBY site. After applying corrections to address the smear effect, the spectral signatures demonstrate improved stability, as shown in the same figure. These corrections significantly enhance the reliability of the data, ensuring consistent spectral characteristics over time, and act as a consistent input to further processing such as vicarious calibration.

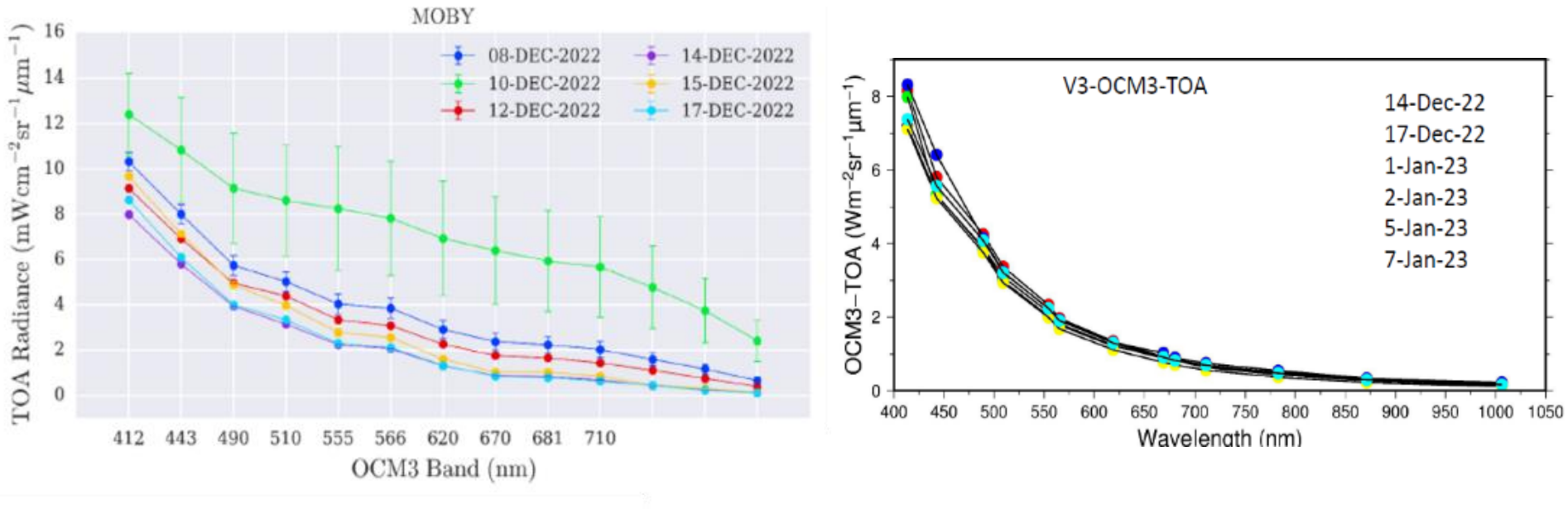

**Figure 6.** MOBY Site Spectras Before and After Smear Correction

**Ground TDI:** In OCM-3, thanks to the imaging scheme, there are multiple samples of same ground feature available on the ground for binning and improving the SNR. By design, the number of samples available for binning are either 23/24 in case of LAC and 7/8 in case of GAC. Because the target is being imaged at different times inorbit, its position on the sampling grid of the detector can be different for each instance and there can be sub-pixel shifts between the observations. This can lead to obvious blurring of the data if the samples are binned directly in nearest neighbor approach. Also, interpolating the data so many number of times with standard interpolaters like cubic, spline also leads to artifacts and blurring. Any ground TDI scheme should compensate for this effect to generate a sharp image with single resampling.

The performance of the OCM-3 satellite can be influenced by several critical factors, including attitude biases that impact image quality and data accuracy. Payload alignment biases arise when the satellite's imaging instruments are not perfectly aligned with the intended observation path, leading to distortions in the captured data. Optical distortions further complicate this issue by introducing aberrations in the image, often caused by imperfections in the optical components. Additionally, platform drift and microvibrations can affect the satellite's stability during image acquisition, resulting in unwanted motion that can blur the captured images. Pixel-to-pixel non-uniformity is another significant concern, where variations in the sensitivity of individual pixels lead to inconsistencies in the recorded signals. Finally, port-to-port dark variations, particularly at the edges of the imaging ports, can introduce discrepancies in dark signal readings, complicating the calibration process. Addressing these critical parameters is essential for ensuring the reliability and accuracy of the data obtained from OCM-3. The ground processing carefully models and mitigates these effects to generate high SNR images.

Following equation gives the mathematical formulation of ground TDI.

$$p(i,j)=\sum_{k=1}^{\substack{K=7/8\\ k=23/24}} \sum_{l=i+\Delta s_k-w_x}^{l=i+\Delta s_k+w_x} \sum_{m=i+\Delta p_k-w_y}^{m=i+\Delta p_k+w_y} w_k(l,m) * g_k(l,m)+n_k(l,m)$$

Output Pixel
LAC/GAC Frames
Neighborhood of ground feature corresponding to (i,j)
Optimal Weight for interpolation
Grey level Value in LAC/GAC Frame
Mapping across frames established through precise & calibrated geometric model where (i,j) is the image location of the output to be constructed, k in the index over LAC/GAC frames, $(i + \dot{\Delta} s_k,\ j + \dot{\Delta} p_k)$ denote for the feature occurring at location(i,j), the shift in the image location in kth LAC/GAC frame, $2w_x$ $and$ $2w_y$ denote the extent of neighborhoods to be taken in account, (l,m) is the corresponding location of ground feature occurring at location (i,j) in kth LAC/GAC frame, $w_k(l,m)$ is the corresponding location of ground feature occurring at location (l,m) in kth LAC/GAC frame, $g_k(l,m)$ is the grey level value corresponding to the neighborhood location (l,m) in kth LAC/GAC frame and $n_k(l,m)$ is the noise value corresponding to the neighborhood location (l,m) in kth LAC/GAC frame.

Figure 7 shows the flowchart of the ground TDI data binning scheme followed for inorbit.

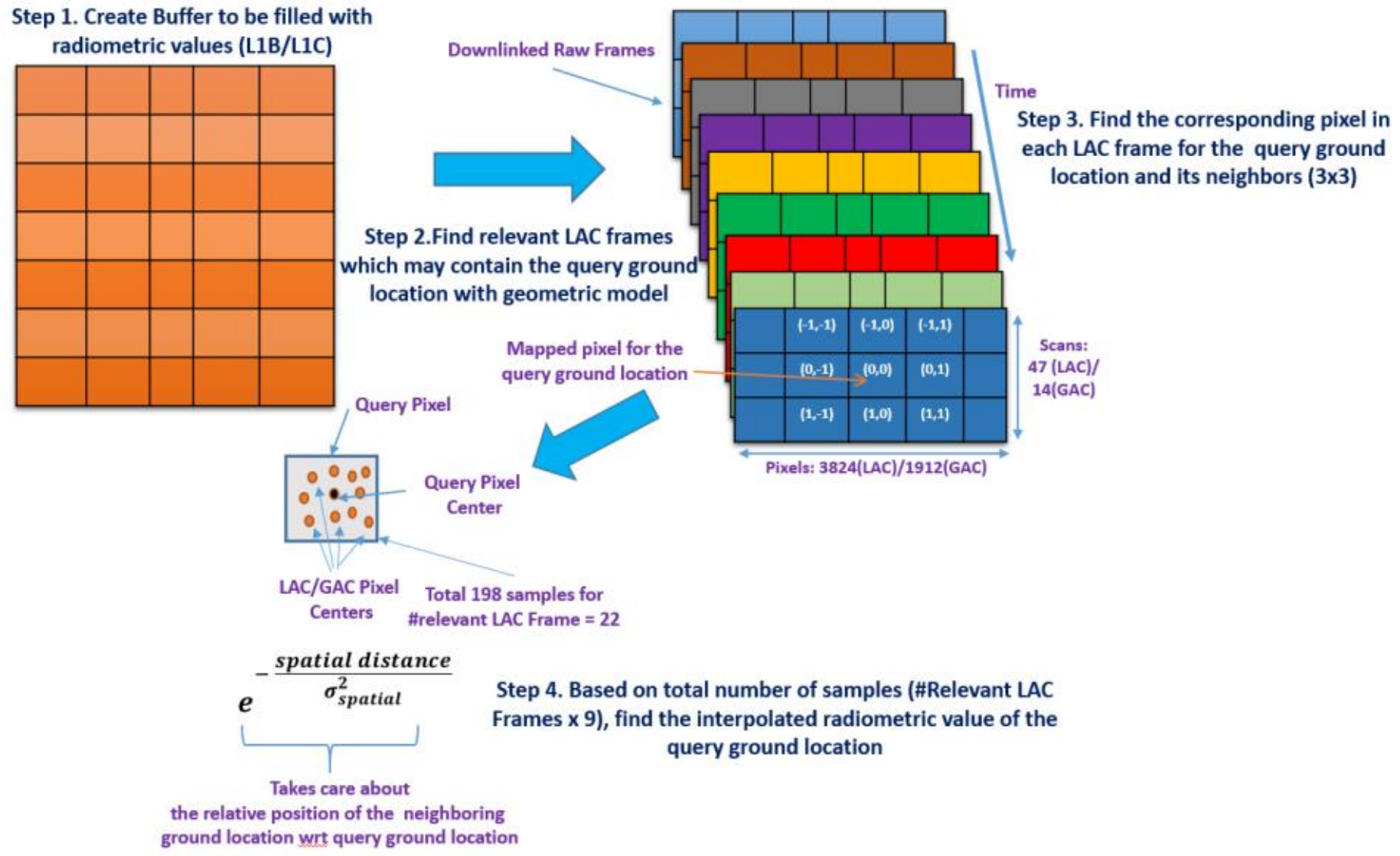

**Figure 7.** Ground Time Delay and Integration Scheme

There are 4 major steps of ground TDI:

**Step-1:** First step comprises of creating an empty buffer to be filled by ground TDI scheme. For L1B, the number of pixels of the empty buffer is equal to the number of pixels in LAC/GAC scheme and number of scan lines is given by the following equation:

num_scans = 47(LAC)/13(GAC)+2*(total raw frames -1)

Using the above calculated number of scan lines and pixels, a geometric model [6] of virtual linear sensor is created having number of pixels equal to LAC/GAC mode and start time equal to the start of first frame. This virtual linear geometric model is very critical and drives the further operations of ground TDI. Also the geometric model for each LAC/GAC frame is calculated for further operations.
**Step-2:** Using the projection parameters for Level 1C or the virtual linear model for Level 1B, the ground location of each pixel in the output buffer is determined. The query ground location is then searched in the relevant LAC/GAC frames.
**Step-3:** Once the relevant frames are identified, the corresponding image locations of the query ground location are calculated, along with their neighboring pixels. The ground locations and radiometric values of these neighbors are stored.
**Step-4:** Final step comprises of find the binned radiometric value based on the samples available. For each pixel calculated in Step-3, the distance between its ground location and the query ground location is calculated. Final radiometric value is given by the following equation:\

$$Binned\ Value = \sum_{i=0}^{\#\ of\ samples} e^{\frac{-distance}{\sigma^2}} * Radiometric\ Value_i$$

The parameter $\sigma$ controls the weight given to each sample based on its distance with respect to the query ground location. A large value of $\sigma$ gives less weightage to neighboring pixels making the scheme behave very similar to nearest neighbor interpolation and can generate unnecessary blurring and artifacts. A small value of $\sigma$ gives more weightage to neighboring pixels and lead to high SNR but smooth image. The value of $\sigma$ is experimentally determined with in orbit data and fixed as 0.5.

Figure 8 shows the comparison between performance achieved by ground TDI scheme with nearest neighbor and sparse interpolation by exponential kernels [7]. It is clearly seen that the second approach leads to better spatial performance and less distortion compared to nearest neighbor approach. The statistical comparison in terms of power spectrum [8] between these two scheme with simulated data from OCM-2 also shows that exponential kernel is able to preserve mid and high frequency information better than nearest neighbor kernel.

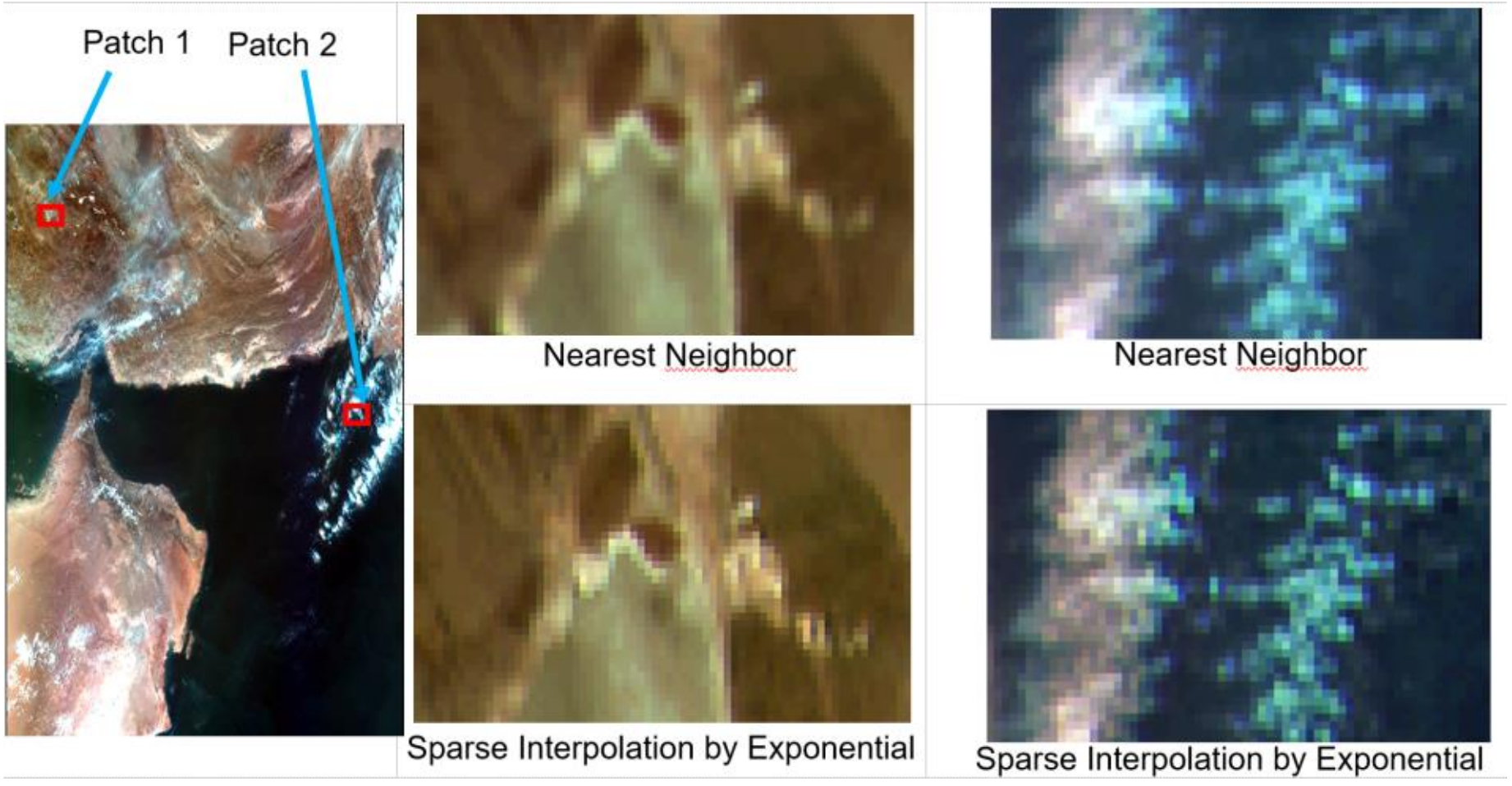

**Figure 8.** Comparison Between Nearest Neighbor and Sparse Interpolation in Ground TDI

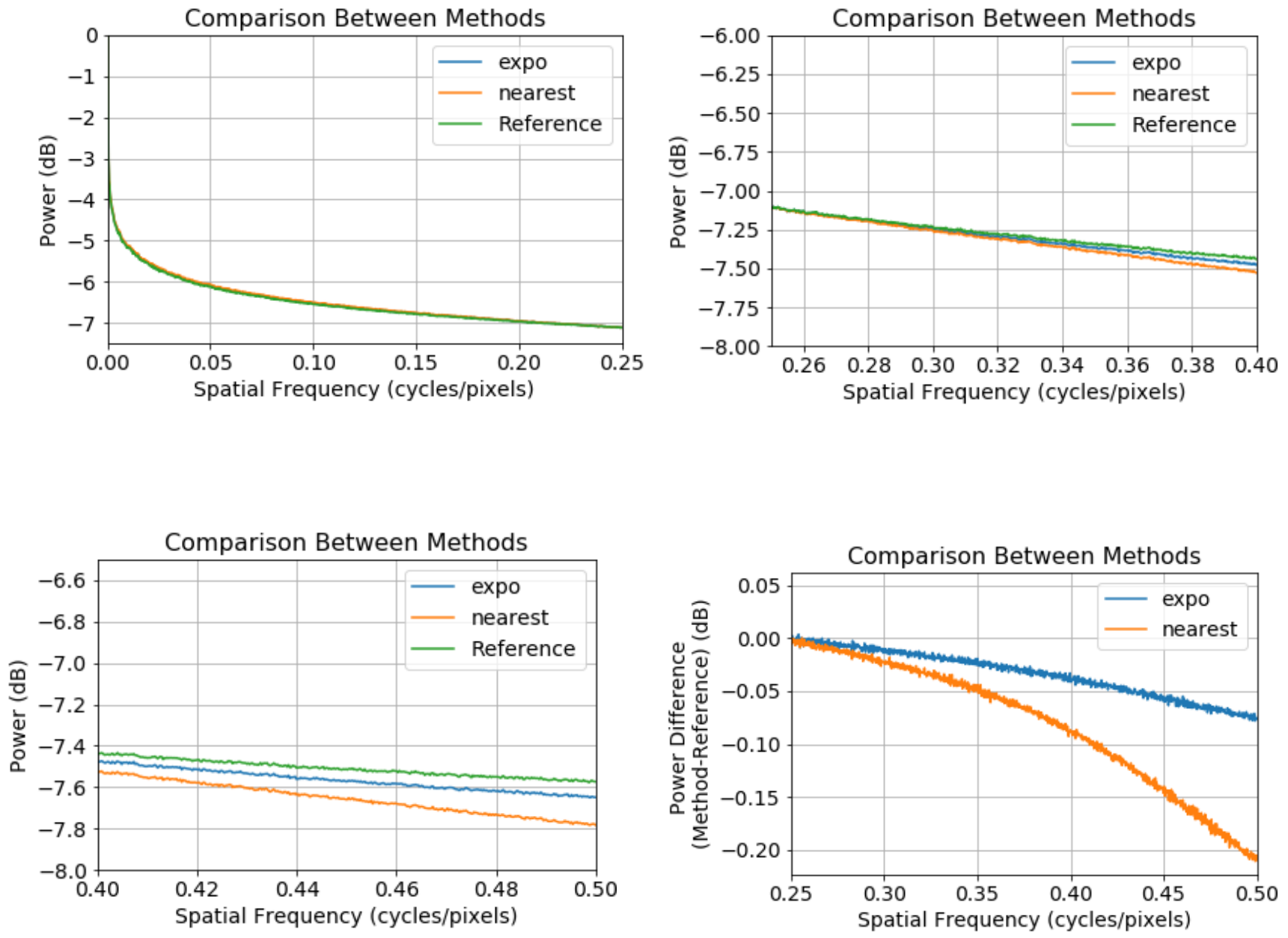

**Figure 9.** Power Spectrum Comparison Between Nearest Neighbor and Sparse Interpolation

Figure 10 shows the number of samples actually available for ground TDI in LAC/GAC mode of acquisition with and without payload tilt. The number of samples available for binning varies with spatial location of the array and can take values from 18-23 for LAC and 5-7 for GAC.

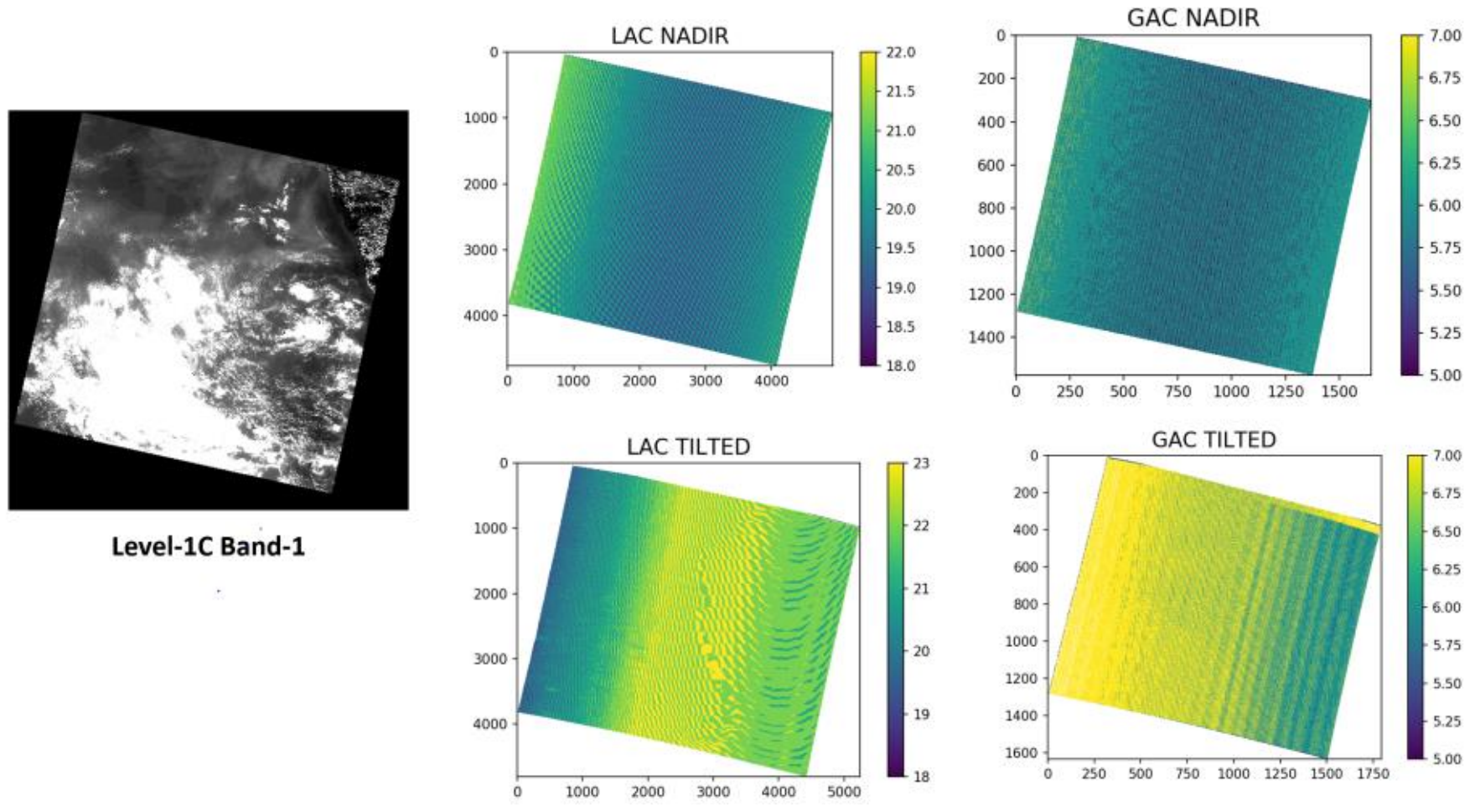

**Figure 10.** Samples Available for LAC and GAC Binning in Nadir and Tilted Configuration

The signal-to-noise ratio (SNR) performance of OCM-3 achieved via ground TDI scheme was thoroughly evaluated in both LAC and GAC modes at the product level. The results indicate that the SNR specifications of 650 (in B1-B10) and 800 (B11-B13) for LAC mode not only met the required thresholds but did so with significant margins, demonstrating the reliability of the system in capturing clear and high-quality data as shown in Figure 11. The evaluation focused on the Level-1C (L1C) product level, which represents the processed data ready for analysis. It was noteworthy that the measured SNR values were higher than initially expected. This enhancement can be attributed to the effective image de-noising techniques implemented within the data processing chain [9]. These techniques play a crucial role in mitigating noise, thereby improving the clarity and accuracy of the images.

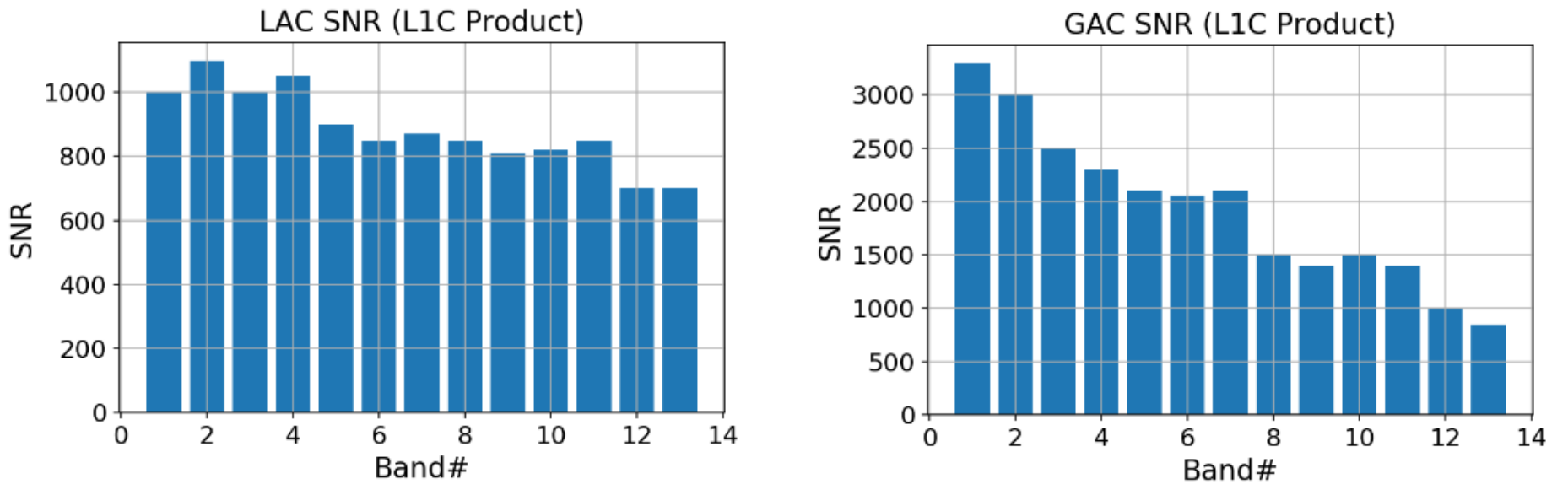

**Figure 11.** Signal to Noise Ratio Achieved at Level-1C

## 4 Geometric Calibration and Performance

Geometric calibration of satellite images is crucial for ensuring the accuracy and reliability of remotely sensed data. This process involves correcting systematic errors that may arise due to various factors, such as sensor misalignment, platform drift, and optical distortions. By addressing these issues, geometric calibration significantly enhances several key performance aspects, including band-to-band registration (BBR), geolocation accuracy, and multi-temporal analysis.

**Band to Band Registration:** Band-to-band misregistration (BBR) for all 12 bands of the OCM-3 satellite was derived from in-orbit data relative to band 7. The methodology employed dense matching techniques to generate tie points between the reference channel and the channels under evaluation [9]. Deviations in both the scan and row directions were subsequently quantified. These deviations were converted into roll and pitch angles, which were utilized in data processing. Prior to correction, the system-level BBR error exhibited variability across the field of view, reaching up to 2 pixels in both the scan and pixel directions due to optical distortions. After applying the corrective measures, the BBR errors were reduced to within the specification of 0.25 pixels as shown in Figure 12 for Band-1 for representation.

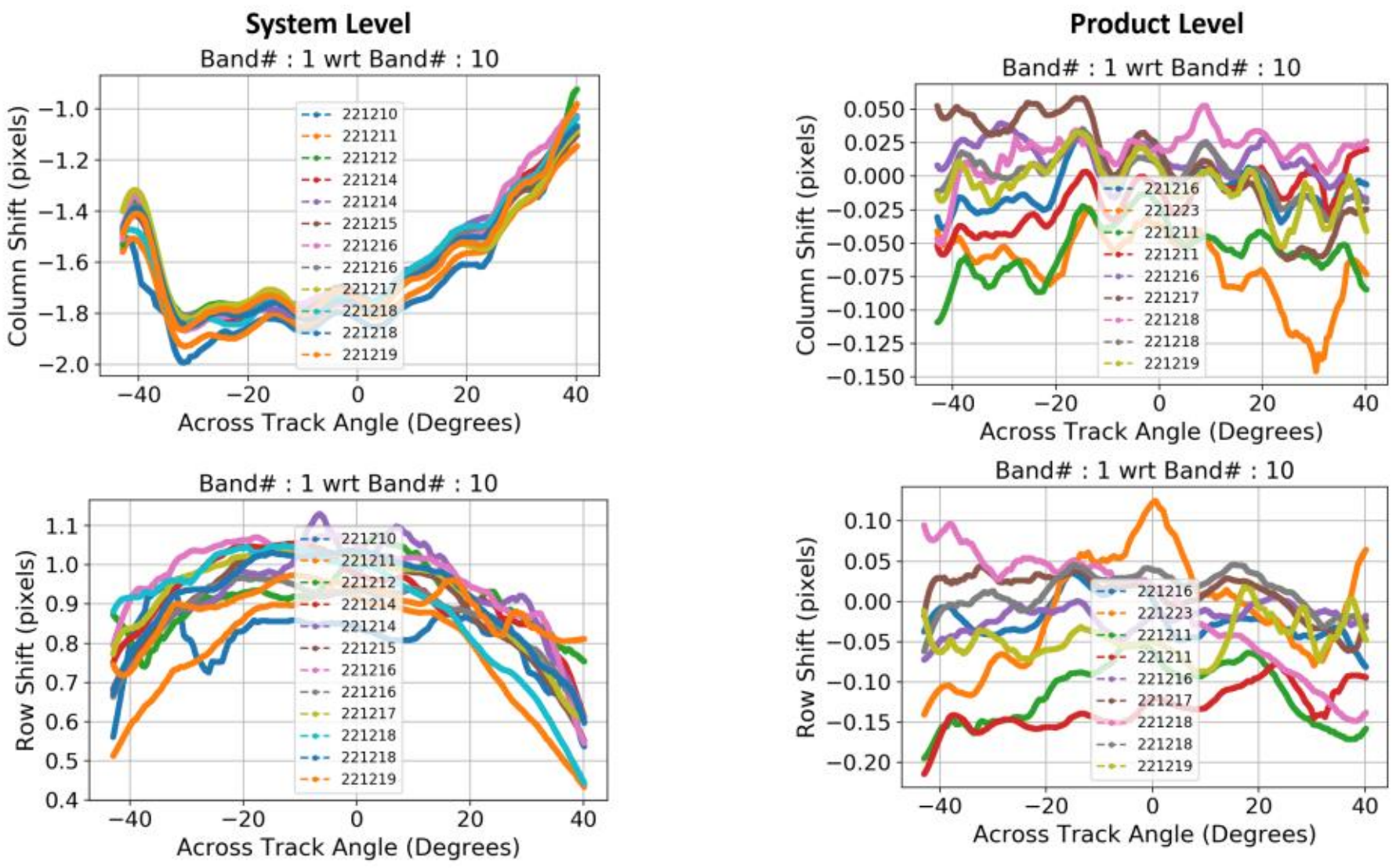

**Figure 12.** Band-1 System Level and Product Label BBR Profile

Table 6 presents the performance metrics achieved, illustrating the efficacy of the calibration processes in mitigating band-to-band misregistration and enhancing data accuracy.

**Table 6.** Band To Band Misregistration Performance After Correction

| Band # | Along Track BBR (pixels) (Mean±3*sigma) | Across Track BBR (pixels) (Mean±3*sigma) |
|---|---|---|
| 1 | 0.082 ± 0.194 | 0.066 ± 0.183 |
| 2 | 0.073 ± 0.180 | 0.060 ± 0.170 |
| 3 | 0.070 ± 0.184 | 0.057 ± 0.167 |
| 4 | 0.065 ± 0.178 | 0.054 ± 0.161 |
| 5 | 0.056 ± 0.169 | 0.048 ± 0.149 |
| 6 | 0.055 ± 0.167 | 0.046 ± 0.145 |
| 7 | 0.048 ± 0.162 | 0.014 ± 0.141 |
| 8 | 0.045 ± 0.169 | 0.040 ± 0.141 |
| 9 | 0.046 ± 0.162 | 0.041 ± 0.135 |
| 11 | 0.017 ± 0.173 | 0.072 ± 0.186 |
| 12 | 0.018 ± 0.175 | 0.083 ± 0.233 |
| 13 | 0.027 ± 0.185 | 0.087 ± 0.228 |

**Geolocation Error:** For the assessment and enhancement of geolocation accuracy, automatic Ground Control Points (GCPs) were identified between the reference channel (Band 10) of the OCM-3 satellite and the Sentinel-3 OLCI reference image. The process of geolocation calibration in OCM-3 is shown in Figure 13.

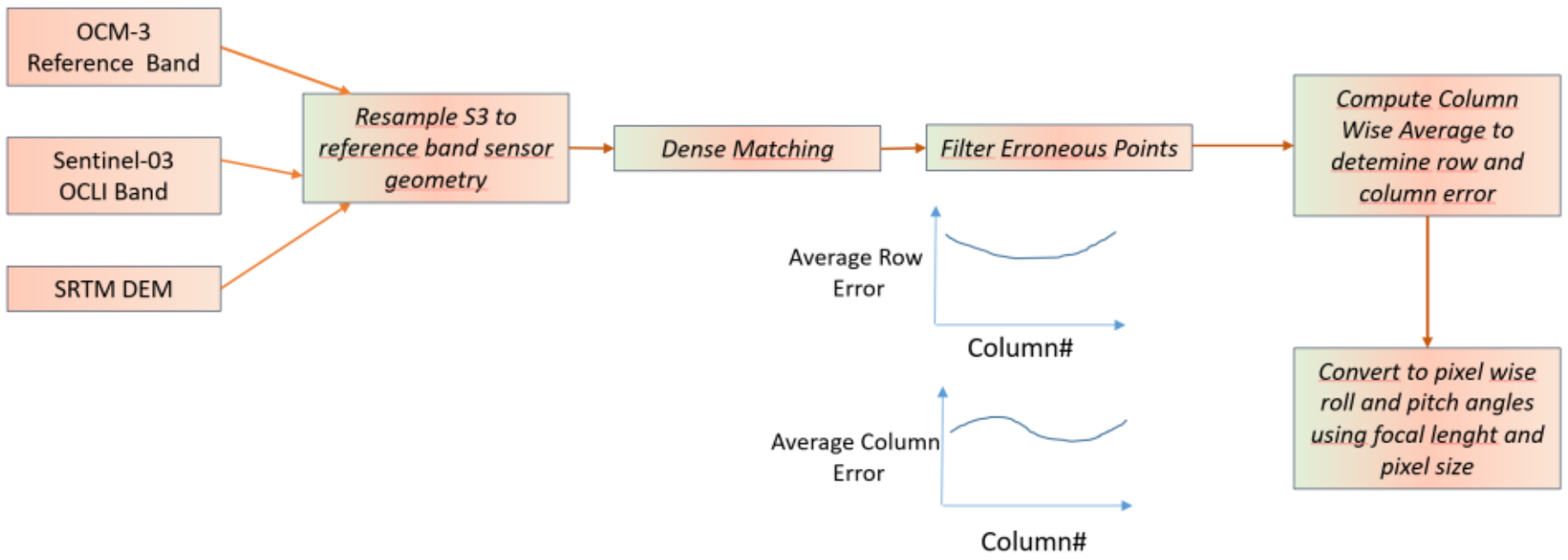

**Figure 13**. Geolocation Calibration Process

This process involved utilizing advanced image registration techniques, enabling the extraction of precise GCPs through dense matching algorithms. By correlating the spectral signatures and spatial characteristics between the two datasets, reliable tie points were established. Residual patterns were generated by analyzing the discrepancies between the matched points in the OCM-3 imagery and the reference data [10,11]. These residuals, which represented the geolocation errors, were systematically examined across the entire field of view as shown in Figure 14. The errors were subsequently converted into pitch and roll angles to facilitate correction, ensuring that the satellite's orientation was accurately accounted for during data acquisition. Table 7 illustrates the error metrics prior to correction, where consistent error patterns were observed across different regions, indicative of systematic biases potentially caused by optical distortions and sensor misalignment. The median error is around 2.5 km in along track direction and around 9 km in across track direction. These patterns highlight the necessity for calibration to achieve optimal geolocation accuracy. Following the application of correction algorithms, which included adjustments in payload alignment angle and interior calibration of the camera based on the derived pitch and roll angles, significant improvements were achieved. Table 8 presents the revised error metrics post-correction, showcasing the reduction in discrepancies. The median error is better than 150m in both across track and along track directions.

**Table 7.** Geolocation Error Before Geometric Calibration

| DOP | Acquisition Mode | Path | Row | Along Track Error Across Swath Variation UB (km) | Along Track Error Across Swath Variation LB (km) | Median Along Track Error (km) | Across Track Across Swath Error Variation UB (km) | Across Track Across Swath Error Variation LB (km) | Median Across Track Error (km) |
|---|---|---|---|---|---|---|---|---|---|
| 221210 | LAC | 51 | 13 | 3.207 | 1.701 | **2.474** | 10.509 | 7.660 | **9.144** |
| 221211 | LAC | 57 | 12 | 3.224 | 1.772 | **2.466** | 10.667 | 7.611 | **9.394** |
| 221211 | LAC | 57 | 13 | 3.252 | 1.696 | **2.490** | 11.903 | 7.725 | **9.254** |
| 221212 | LAC | 50 | 13 | 3.211 | 1.677 | **2.426** | 10.361 | 7.529 | **9.332** |
| 221212 | LAC | 64 | 14 | 3.265 | 1.666 | **2.574** | 11.726 | 7.065 | **8.813** |
| 221214 | LAC | 48 | 12 | 3.370 | 1.666 | **2.442** | 11.156 | 7.727 | **9.457** |
| 221214 | LAC | 63 | 14 | 3.206 | 1.734 | **2.576** | 10.675 | 6.557 | **9.012** |
| 221215 | LAC | 54 | 13 | 3.247 | 1.799 | **2.527** | 11.698 | 6.909 | **9.210** |
| 221216 | LAC | 48 | 14 | 3.343 | 1.665 | **2.545** | 10.909 | 7.206 | **9.190** |
| 221217 | LAC | 53 | 12 | 3.213 | 1.664 | **2.455** | 10.868 | 7.160 | **9.288** |
| 221217 | LAC | 53 | 13 | 3.223 | 1.705 | **2.515** | 10.477 | 7.197 | **8.854** |
| 221218 | LAC | 58 | 12 | 3.267 | 1.718 | **2.531** | 11.028 | 7.201 | **9.230** |
| 221218 | LAC | 59 | 13 | 3.261 | 1.718 | **2.540** | 11.335 | 6.823 | **9.158** |
| 221219 | LAC | 66 | 14 | 3.310 | 1.777 | **2.562** | 11.496 | 7.174 | **8.671** |

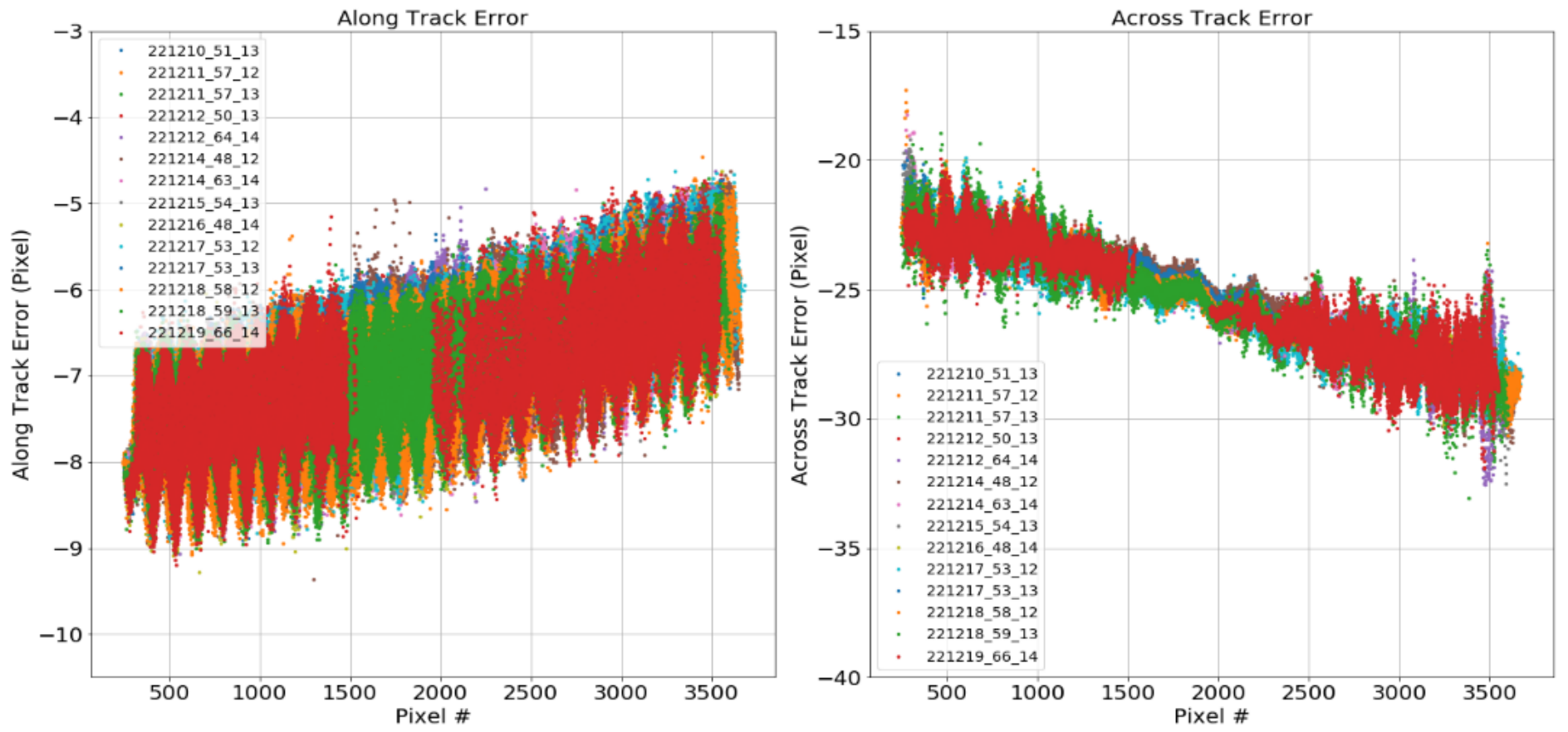

**Figure 14.** Residual Along and Across Track Error Before Geometric Calibration

Another phenomena in OCM-3 which affected geolocation accuracy was related to tilt angle of the payload. The OCM-3 satellite utilizes a payload tilt mechanism to mitigate sun glint effects during image acquisition. Observations revealed a continuous variation in payload tilt angles throughout the imaging orbit. This variation had a significant impact on geolocation accuracy, characterized by a progressive increase in the along-track mean error correlated with the tilt angle. In contrast, the across-track mean errors remained relatively consistent. Figure 15 illustrates these observations during the initial operational days of the satellite. To quantify the relationship between the tilt angle and the resulting errors, a linear model was established correlating pitch residuals with changes in tilt angle. By modeling this drift and incorporating the corrected tilt angle in the pitch direction, the calibration process effectively addressed the induced errors as shown in Figure 16.

**Table 8.** Geolocation Error After Geometric Calibration

| **DOP** | **Acquisition Mode** | **Path** | **Row** | **Along Track Error Across Swath Variation UB (km)** | **Along Track Error Across Swath Variation LB (km)** | **Median Along Track Error (km)** | **Across Track Across Swath Error Variation UB (km)** | **Across Track Across Swath Error Variation LB (km)** | **Median Across Track Error (km)** |
|---|---|---|---|---|---|---|---|---|---|
| 221210 | LAC | 51 | 13 | 0.222 | 0.510 | **0.114** | 0.262 | 0.318 | **0.004** |
| 221211 | LAC | 57 | 12 | 0.206 | 0.252 | **0.090** | 0.128 | 0.238 | **0.094** |
| 221211 | LAC | 57 | 13 | 0.219 | 0.468 | **0.109** | 0.209 | 0.324 | **0.027** |
| 221212 | LAC | 50 | 13 | 0.164 | 0.473 | **0.151** | 0.241 | 0.269 | **0.012** |
| 221212 | LAC | 64 | 14 | 0.277 | 0.469 | **0.140** | 0.212 | 0.298 | **0.078** |
| 221214 | LAC | 48 | 12 | 0.271 | 0.342 | **0.077** | 0.242 | 0.289 | **0.128** |
| 221214 | LAC | 63 | 14 | 0.401 | 0.417 | **0.103** | 0.216 | 0.227 | **0.096** |
| 221215 | LAC | 54 | 13 | 0.280 | 0.422 | **0.077** | 0.104 | 0.291 | **0.029** |
| 221216 | LAC | 48 | 14 | 0.145 | 0.353 | **0.083** | 0.235 | 0.272 | **0.049** |
| 221217 | LAC | 53 | 12 | 0.125 | 0.393 | **0.111** | 0.129 | 0.224 | **0.145** |
| 221217 | LAC | 53 | 13 | 0.202 | 0.463 | **0.127** | 0.187 | 0.279 | **0.022** |
| 221218 | LAC | 58 | 12 | 0.146 | 0.300 | **0.074** | 0.202 | 0.271 | **0.133** |
| 221218 | LAC | 59 | 13 | 0.124 | 0.282 | **0.079** | 0.175 | 0.231 | **0.021** |
| 221219 | LAC | 66 | 14 | 0.131 | 0.323 | **0.059** | 0.219 | 0.237 | **0.069** |

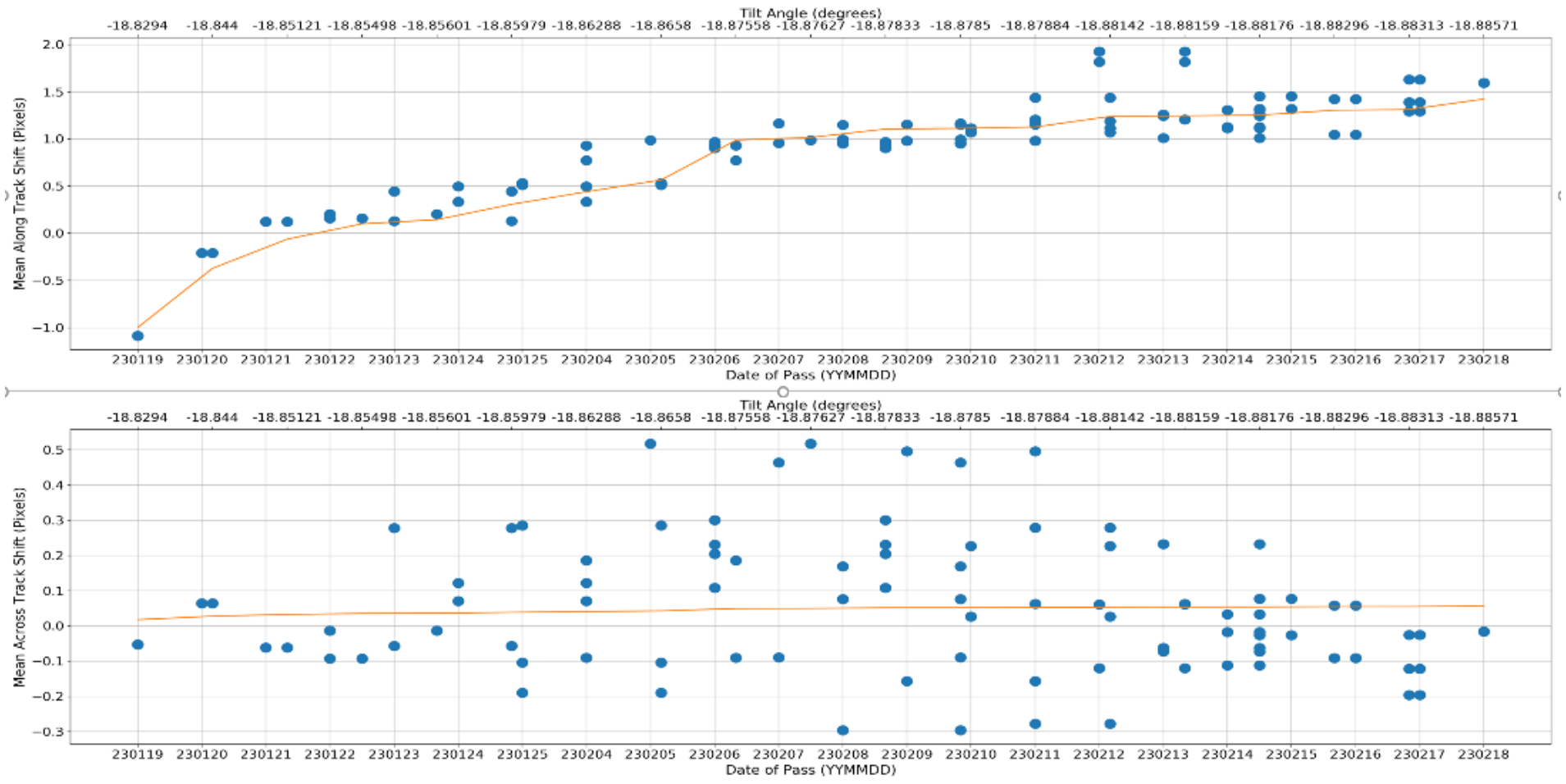

**Figure 15.** Tilt Related Geometric Error

Post-correction, the accuracy of the geolocation data improved markedly, with both along-track and across-track errors reduced to within the specified threshold of 0.25 pixels. This successful calibration underscores the effectiveness of the tilt mechanism and correction methodologies in enhancing the overall data quality and precision of the OCM-3 satellite.

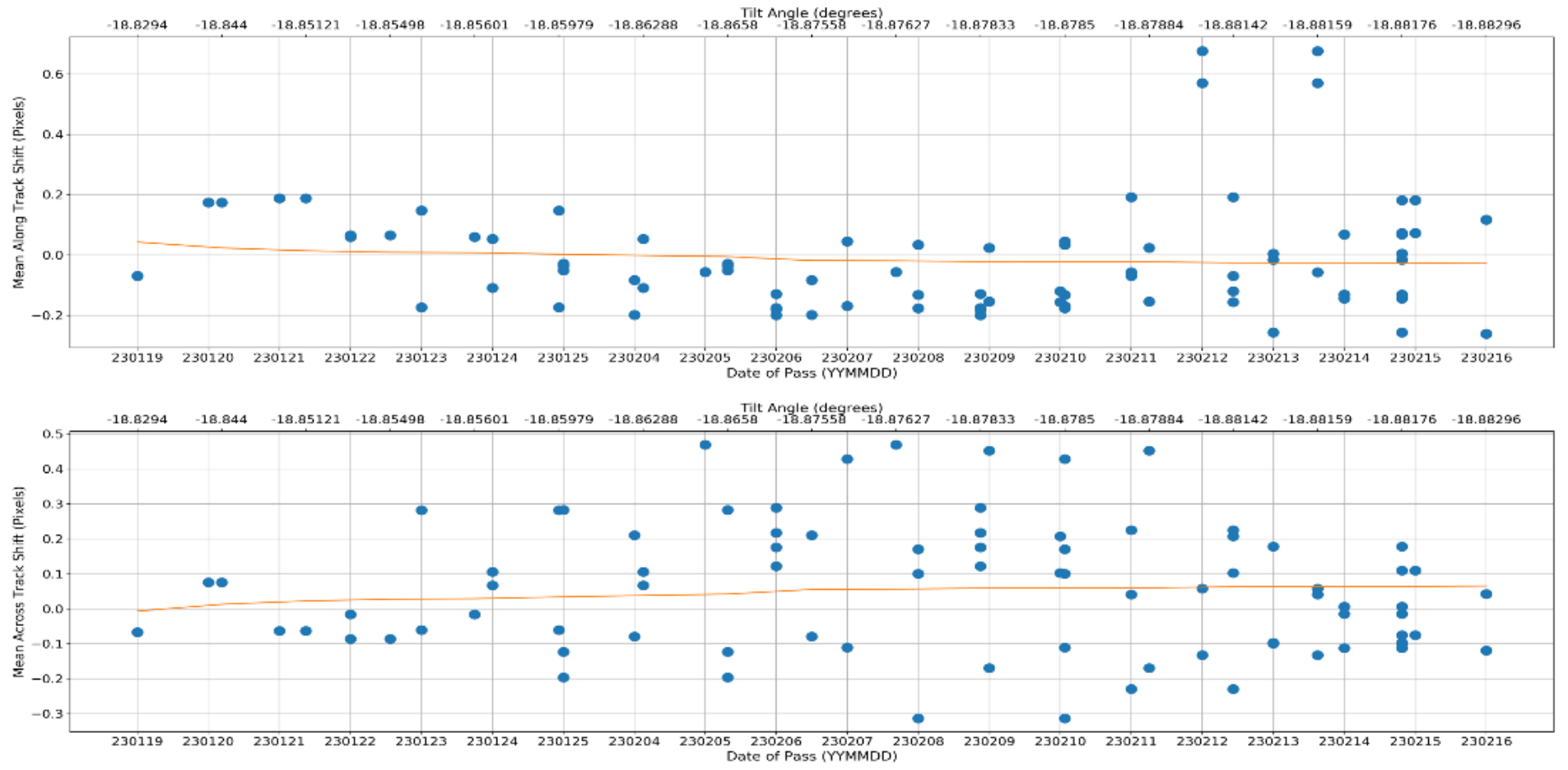

**Figure 16.** Error After Tilt Related Geometric Error

Figure 17 depicts the two-dimensional error pattern after calibration generated across large number of scenes from around six months of data from Feb 2024 to September 2024, revealing that the errors now fall within the specified limits for both LAC and GAC mode of aquisition. It shows that the requirement of this parameter is met with very good margins. This successful calibration demonstrates the efficacy of the tech-niques employed in mitigating geolocation errors and enhancing the overall data quality of the OCM-3 satellite.

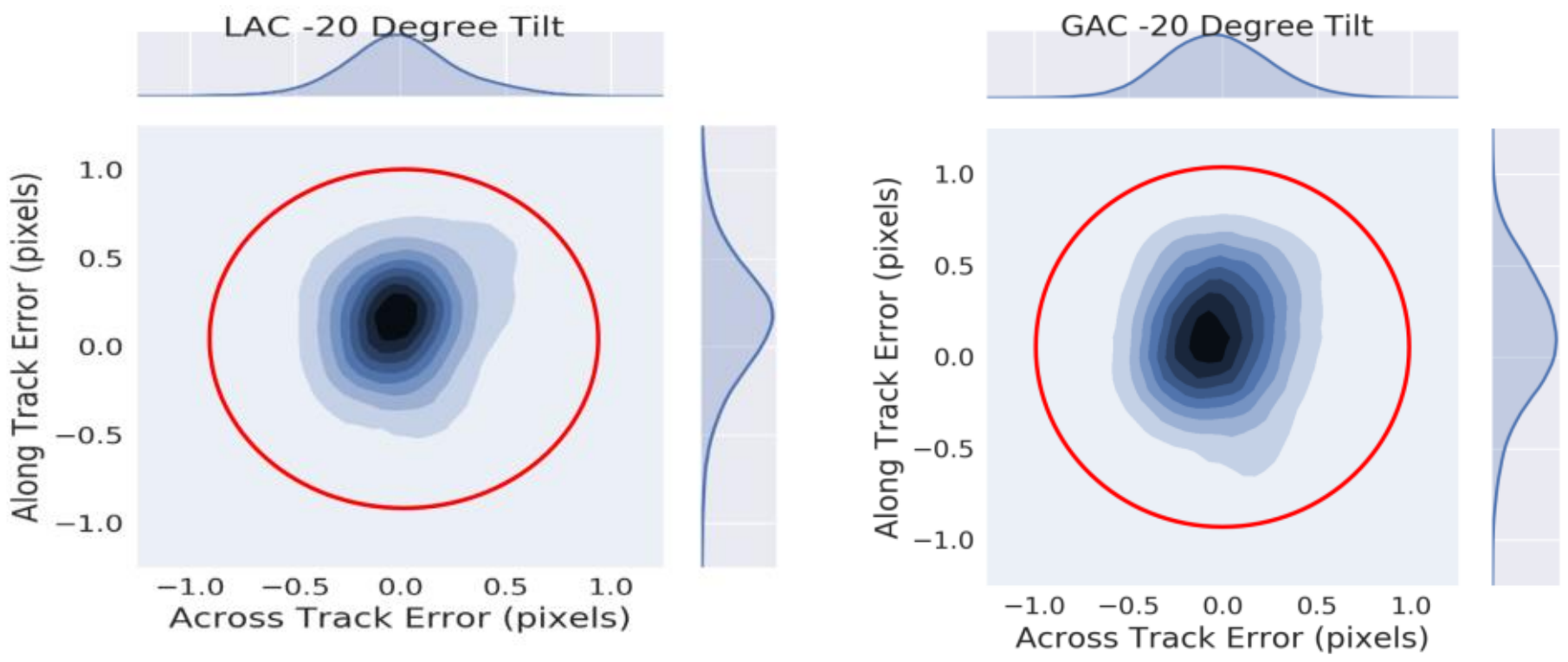

**Figure 17.** Geolocation Error After Geometric Calibration

**Multi-temporal Accuracy:** Multi-temporal registration is essential in remote sensing to accurately align images acquired at different times, enabling applications such as change detection and land use monitoring. Key requirements for effective registration include geometric consistency, which ensures that corresponding features align accurately across time periods. It is very critical design parameter for change detection analysis [12]. Figure 18 shows the multi-temporal accuracy achieved after geometric calibration across large number of scenes from around six months of data from Feb 2024 to September 2024. It shows that the requirement of this parameter is met with very good margins.

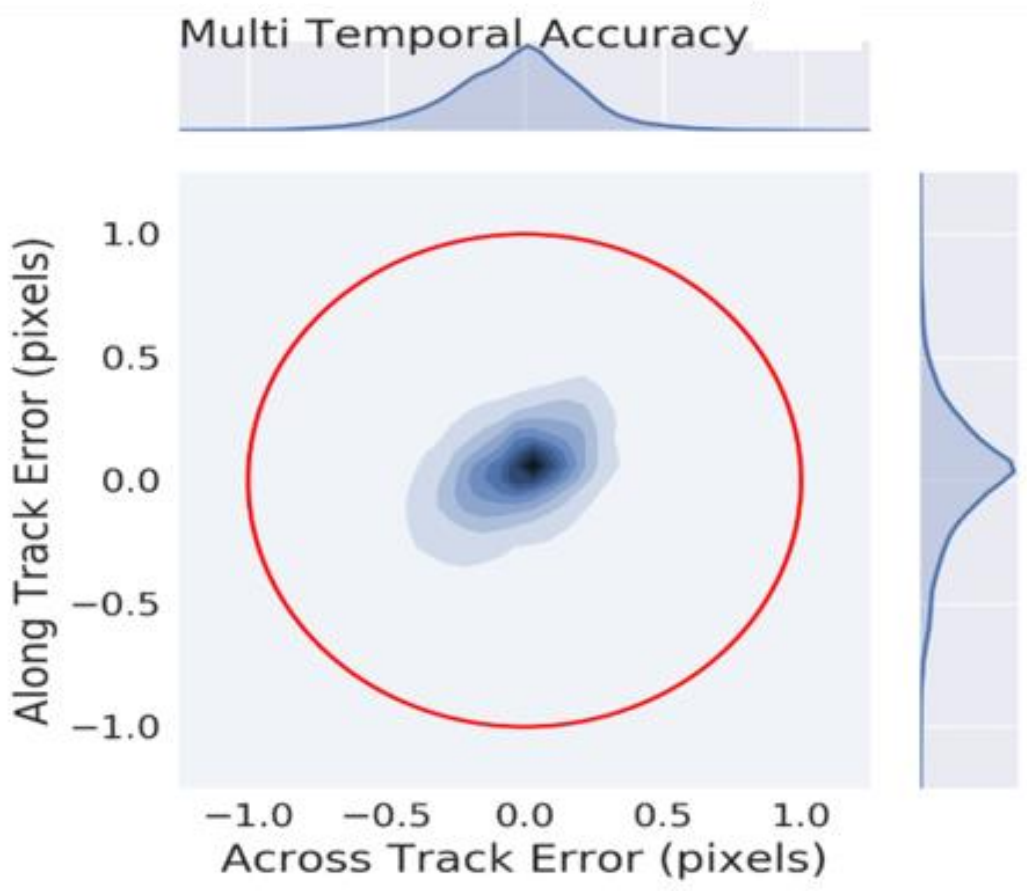

**Figure 18.** Multitemporal Accuracy After Geometric Calibration

# 5 OCM-3 Images

The initial operational capabilities of the Ocean Color Monitor-3 (OCM-3) are exemplified in the imagery captured on its first day of operation. Figure 19 showcases both natural and false color composite images, highlighting the satellite's ability to distinguish between different water types and land features through enhanced spectral resolution.

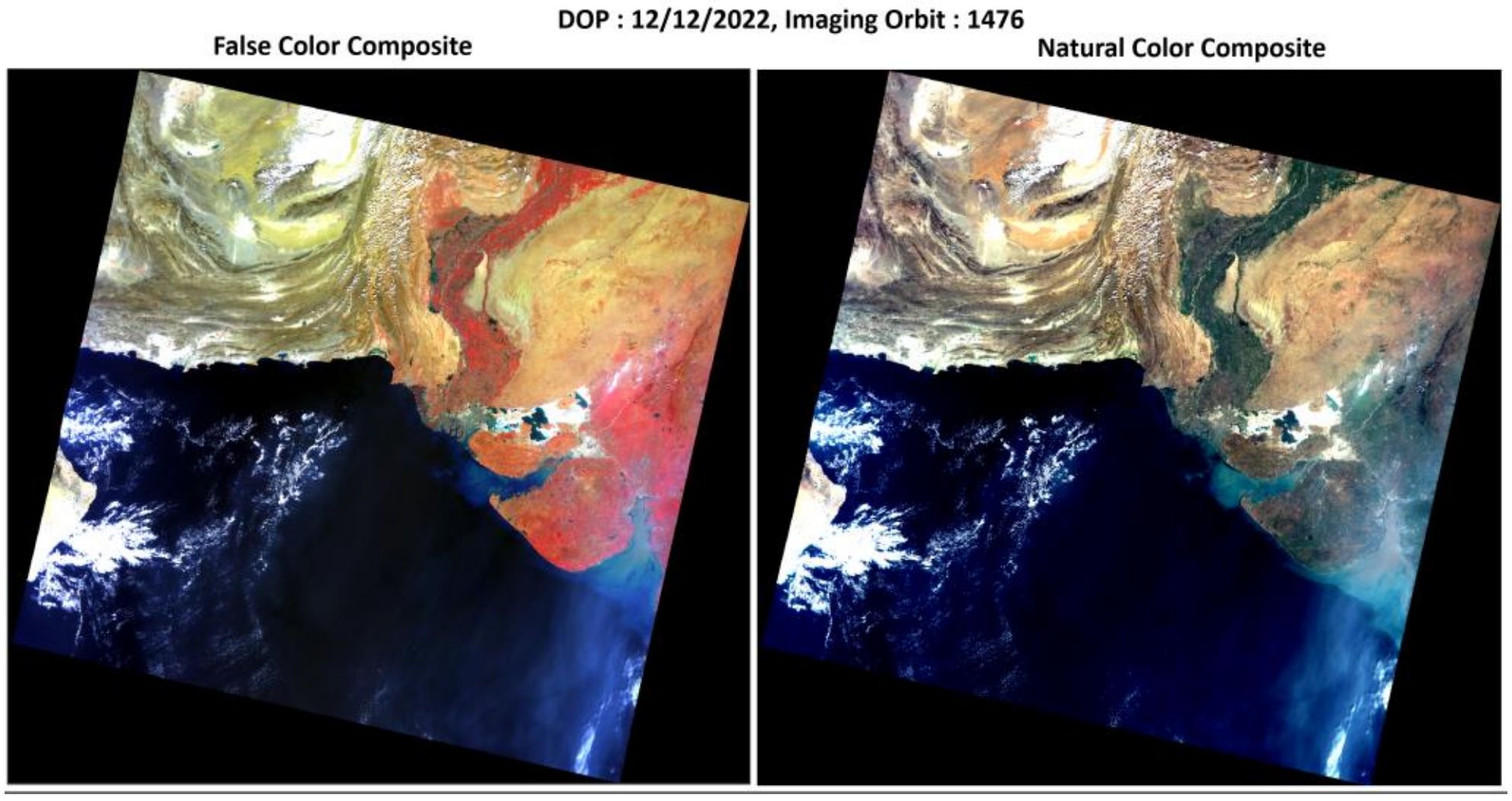

**Figure 19.** OCM-3 First Day Natural and False Color Composite Image

In Figure 20, a global mosaic generated from OCM-3's Global Area Coverage (GAC) mode demonstrates the satellite's extensive observational reach, providing a comprehensive view of oceanic and terrestrial environments. This mosaic is instrumental in assessing large-scale environmental changes and monitoring ocean color dynamics across vast regions.

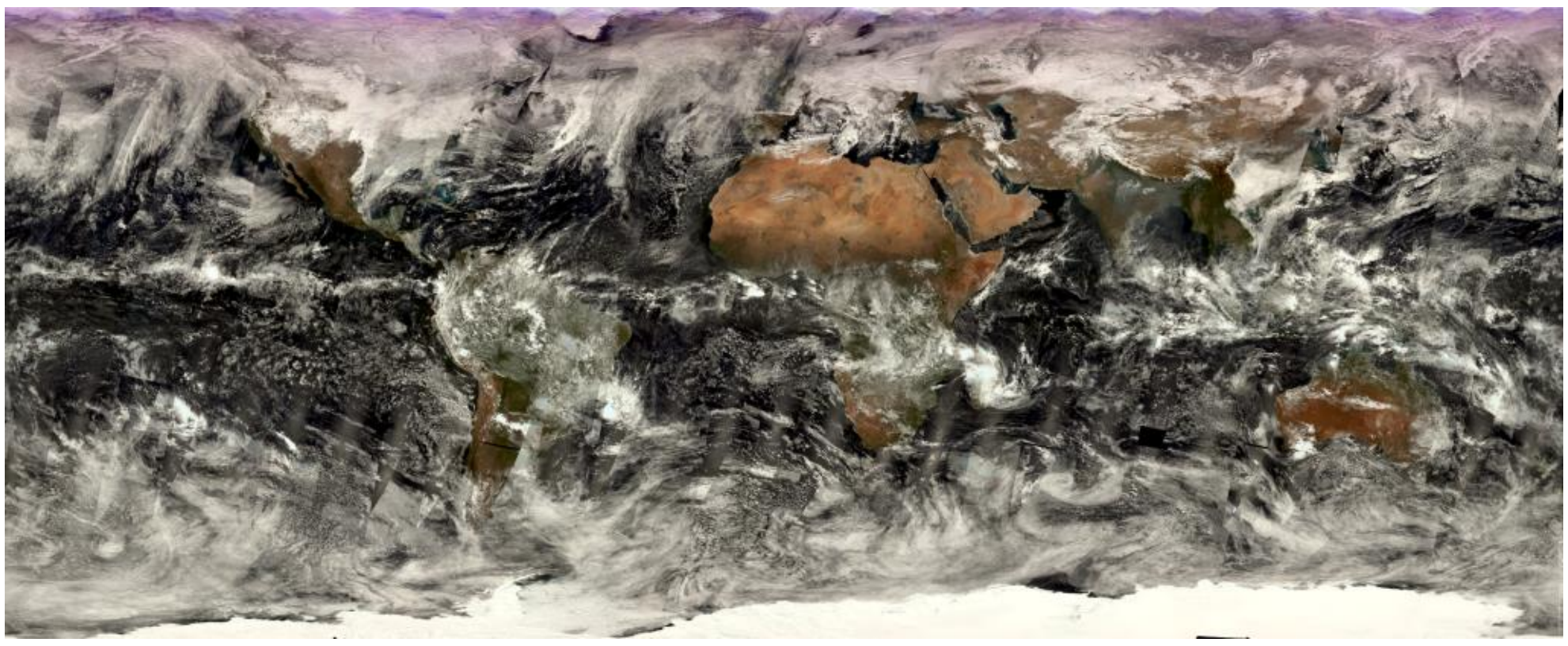

**Figure 20.** Global Mosaic From OCM-3 GAC Mode

Furthermore, Figure 21 presents a detailed mosaic over Antarctica, specifically showcasing the Maitri and Bharati Research Stations. This image underscores OCM-3's capacity to support scientific research in polar regions, offering vital data for understanding climatic and ecological changes in these sensitive environments. Together, these figures illustrate the diverse applications and significant contributions of OCM-3 to Earth observation.

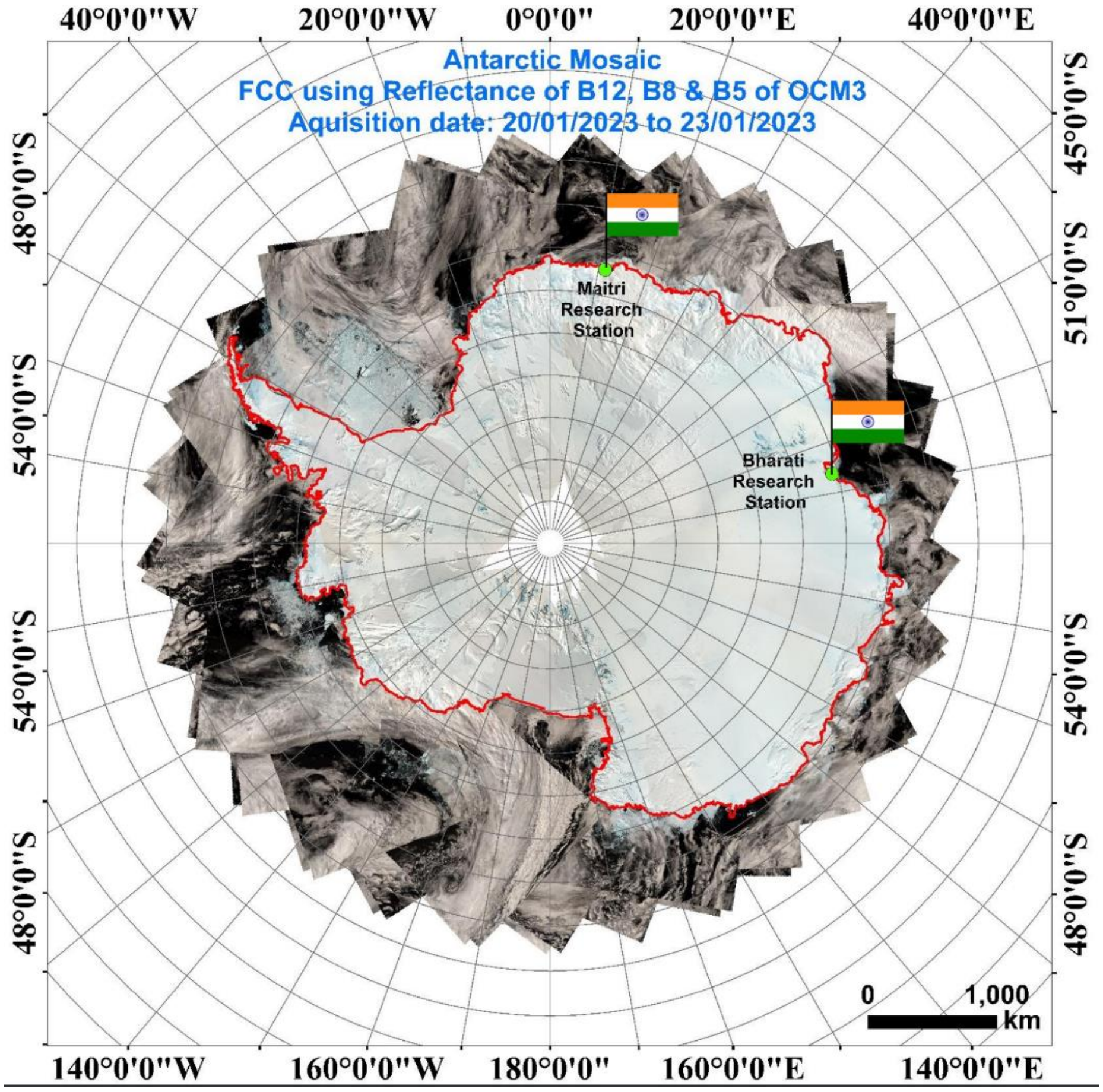

**Figure 21.** OCM-3 Mosaic Over Antarctica showing Maitri and Bharati Research Station

## Conclusion

The Ocean Color Monitor-3 (OCM-3), launched on the Oceansat-3 satellite, advances ocean monitoring through its thirteen spectral bands, offering improved data quality compared to earlier models. Operating in a sun-synchronous orbit, OCM-3 captures high-resolution imagery with enhanced signal-to-noise ratios (SNR) achieved via sophisticated processing techniques. Key methods such as dark data modeling, photo response non-uniformity correction, and smear correction contribute to the accuracy and reliability of the collected data.

Geometric calibration plays a critical role in OCM-3's effectiveness, addressing band-to-band registration (BBR) and geolocation errors. The BBR is maintained within ±0.25 pixels, while advanced image registration techniques enhance geolocation accuracy. Additionally, a payload tilt mechanism is employed to reduce sun glint effects, ensuring precise spatial representation. The satellite's ability to align images over time supports important applications like change detection and land use monitoring.

Overall, OCM-3's advanced calibration processes and high-resolution capabilities make it an essential tool for environmental monitoring and research, improving our understanding of ocean dynamics and informing policy decisions.